\def\year{2022}\relax
\documentclass[letterpaper]{article} 
\usepackage{aaai22}  
\usepackage{times}  
\usepackage{helvet}  
\usepackage{courier}  
\usepackage[hyphens]{url}  
\usepackage{graphicx} 
\usepackage{natbib}  
\usepackage{caption} 
\DeclareCaptionStyle{ruled}{labelfont=normalfont,labelsep=colon,strut=off} 
\usepackage{algorithm}
\usepackage{algorithmic}

\usepackage{newfloat}
\usepackage{listings}
\floatstyle{ruled}
\newfloat{listing}{tb}{lst}{}
\floatname{listing}{Listing}
\usepackage{makecell}
\usepackage[dvipsnames]{xcolor}
\usepackage{footnote}
\usepackage{footmisc}
\usepackage{amsmath}
\usepackage{multirow}
\usepackage{threeparttable}

\makesavenoteenv{tabular} 
\makesavenoteenv{table}

\newcommand{\eg}{{e.g.,\ }}
\newcommand{\etal}{{et al.\ }}

\newcommand{\ie}{{i.e.,\ }}

    \title{Know It to Defeat It: Exploring Health Rumor Characteristics and Debunking Efforts on Chinese Social Media during COVID-19 Crisis}

\author{
    Wenjie Yang,\textsuperscript{\rm 1}
    Sitong Wang,\equalcontrib\textsuperscript{\rm 2}\thanks{This work was done while the author was at HKUST.}
    Zhenhui Peng,\equalcontrib\textsuperscript{\rm 1}
    Chuhan Shi,\textsuperscript{\rm 1}
    Xiaojuan Ma,\textsuperscript{\rm 1}
    Diyi Yang\textsuperscript{\rm 3}
    \\
}

\affiliations{
    \textsuperscript{\rm 1}The Hong Kong University of Science and Technology\\
    \textsuperscript{\rm 2}Columbia University\\
    \textsuperscript{\rm 3}Georgia Institute of Technology \\
    \{wyangbc,zpengab,cshiag\}@connect.ust.hk,
    sw3504@columbia.edu,
    mxj@cse.ust.hk,
    diyi.yang@cc.gatech.edu

}

\begin{document}

\maketitle

%
%
%

\begin{abstract}
Health-related rumors being spread online during a public crisis may pose a serious threat to people's well-being. %
Existing crisis informatics research lacks in-depth insights into the characteristics of 
\textit{health rumors} and \textit{the efforts to debunk them} on social media in a pandemic. To fill this gap,
we conduct a comprehensive analysis of four months of rumor-related online discussion during COVID-19 on Weibo,
a Chinese microblogging site. 
Results suggest that the dread (cause fear) type of health rumors provoked significantly more discussions and lasted longer than the wish (raise hope) type.
We further explore how four kinds of social media users (\ie government, media, organization, and individual) combat health rumors, and identify their preferred way of sharing the debunking information and the key rhetoric strategies  used in the process.
We examine the relationship between debunking and rumor discussions using a Granger causality approach, and show the efficacy of debunking in suppressing rumor discussions, which is time-sensitive and varies according to rumor type and debunker.
Our results can provide insights into crisis informatics and risk management on social media in pandemic settings.
\end{abstract}

\section{Introduction}

During the COVID-19 pandemic, an overabundance of rumors spread broadly on social media \cite{brennen2020types, islam2020covid}. In particular, \textit{health rumors} -- unverified information concerning the practice of healthcare and medicine \cite{cullen2006health} -- can pose a major threat to public health by misleading people into action that is potentially harmful to their well-being \cite{ghenai2017health}. 
Many individuals and groups, either for self-interest or altruistic reasons, participate in countering these rumors as debunkers on social media \cite{brennen2020types}. The outcome of this war between health rumors and debunkers can have a profound impact on decisions regarding health emergencies, risk management, and policy-making during a public health crisis \cite{gui2017understanding}. 
It is hence necessary to gain an in-depth understanding of the characteristics of health rumors and the efficacy of assorted debunking efforts.
Previous research has extensively studied  the rumoring and debunking process on social media in the contexts of short-term 
extreme events \eg natural disasters \cite{oh2010exploration, rajdev2015fake} and human-induced disasters \cite{arif2016information, arif2017closer}. Unlike those events, public health crises could last for a long time, affect a broad geographical scope, and impact a large population \cite{gui2018multidimensional}. Thus, it would require analyzing rumor and counter-rumor efforts on a greater temporal and social scale. 
While recent works have looked into the dissemination of misinformation and conspiracy theories during health crises like Zika virus \cite{ghenai2017catching, kou2017conspiracy} and COVID-19 \cite{brennen2020types, islam2020covid, tasnim2020impact},
there still lacks a systematical investigation into health rumors, their distinctive characteristics, and debunking activities in those situations.
Health rumors are typically categorized into \textit{dread} and \textit{wish} types according to people’s underlying concerns \cite{difonzo2012rumors}. As the name implies, dread rumors invoke fearsome consequences (\eg microwave use can spread cancer), while wish rumors promote favorable outcomes (\eg vitamin can cure cancer). Past research has argued that people tend to pay more attention to dread rumors than to wish rumors 
\cite{difonzo2007rumor}%
, as people are psychologically more  interested in bad news than good news  \cite{baumeister2001bad}.
This conclusion was confirmed by small-scale user experiments with laypeople \cite{difonzo2008watercooler, difonzo2012rumors} and medical professionals \cite{chua2018intentions}. However, the differences between these two types of health rumors on social media have not been well studied, especially in the context of public health crises. Not to mention there is insufficient understanding of debunking efforts against them.
Past research has investigated the active engagement of particular social media user groups, such as journalists \cite{andrews2016keeping} and official organizations \cite{starbird2018engage}, in debunking activities related to extreme events \cite{Chen2020ConspiracyAD}. However, prior studies also pointed out that these debunkers may face dilemmas in health crises. For instance, authorities' perceptions of information during health crises could be highly uncertain \cite{gui2017managing}, leading to their insufficient involvement in risk communication \cite{gui2017understanding}. 
Questions then arise as to what kinds of social media users are actively debunking such rumors during a health crisis, what strategies they tend to apply, and how effective these counter-efforts are. %
To fill the research gaps identified above, this paper aims to address the following questions:
\textbf{RQ1)} What are the characteristics of online health rumors during the COVID-19 crisis, and how do these characteristics differ in terms of dread and wish rumors? 
\textbf{RQ2)} What kinds of social media users have contributed to countering health rumors on social media during the pandemic, and how do the efforts of these debunkers differ? 
\textbf{RQ3)} What are the effects of debunking, and how do the effects vary across types of health rumors and debunkers? 
Knowing the answers to these underexplored questions could help better tackle the infodemic \cite{zarocostas2020fight} in the fight against COVID-19 and other public health crises.
To this end, we carry out a series of visualization and quantitative analysis on 
health rumor discussion and debunking posts from Weibo (\ie a Chinese microblogging site) and various Chinese fact-checking websites. 
Our findings reveal that dread rumors were generally more viral in nature than wish rumors, except for extreme wish rumors. 
We also show that debunking is useful in forecasting and suppressing rumor discussion and that the suppression effect varies across the two rumor types and across the four kinds of debunkers, namely \textit{individuals}, \textit{organizations}, \textit{media}, and \textit{government}.
Based on our findings, we provide both theoretical and practical implications for better understanding of health rumors and rumor debunking. 
We make our annotated data publicly available for future research\footnote{\url{https://github.com/Kelaxon/COVID19-Health-Rumor}}.
%
%
%
%
%

%
%
%
%
%

%
%
%
%
%
%
%
%
%
%
%
%


\section{Related Work}

Health-related misinformation and rumors circulating on social media pose a substantial threat to the public \cite{smailhodzic2016social}, influencing people’s medical decisions and even threatening their lives \cite{wang2019systematic}.
To cope with this issue and understand the nature of health rumors, previous research has extensively investigated the spread of health rumors on diverse health topics, such as the anti-vaccine movement \cite{nyhan2014effective, dredze2016zika} and cancer treatments \cite{chen2018nature, difonzo2012rumors}. Based on early ideas of rumor psychology, Difonzo \etal \cite {difonzo2012rumors} classified cancer rumors into wish and dread categories based on the expected consequences. They collected rumors recalled by users of online cancer communities through questionnaires and concluded that dread rumors outnumbered wish ones. Subsequent studies also found the differences between the two types of health rumors in terms of their psychological impacts on individuals \cite{chua2017share, chua2018intentions}. These findings reflect the classic psychological view that ``bad is stronger than good'' \cite{cacioppo1994relationship, baumeister2001bad}. However, these findings have not been further validated by empirical studies on social media.

In recent years, informatics in public health crises has received much scholarly attention. A health crisis can exacerbate information uncertainty \cite{gui2017managing} and reduce trust towards authorities \cite{freeman2020coronavirus}, very likely leading to the occurrence of conspiracy theories -- another type of misinformation \cite{bruder2013measuring, van2013belief} that is the main focus of many recent crisis informatics studies. %
However, the health rumors we focus on are different and only involve health knowledge, \eg ``vitamins can cure coronavirus''.
The subjects and scenarios of interest in our paper fill a gap in the existing literature and deepen the understanding of health rumor characteristics.

The growing danger of misinformation on social media has prompted research on rumor interventions and debunking. Previous findings about the impact of debunking on misinformation are conflicting, ranging from the ``backfire effect''\cite{nyhan2014effective} to ``effective" \cite{shin2017political}. The effectiveness of debunking may correlate with many factors, \eg the source and content of the rumor \cite{walter2020evaluating}. In the health field, several studies support the effectiveness of debunking \cite{ozturk2015combating, pal2019debunking}. For example, Ozturk \etal find the presence of refutation can decrease the likelihood of health rumor sharing \cite{ozturk2015combating}. However, empirical evidence demonstrating the debunking effect on health rumors across large-scale social networks is limited.

Debunking activities on social networks often take the form of ``wisdom of crowds'' \cite{tanaka2013toward, arif2017closer}, especially when traditional risk communication channels may not be available in a timely manner during crisis events. The identification of information in such cases often relies on self-correction of online crowds \cite{arif2017closer}. However, different types of user groups may have different levels of engagement. For example, journalists on Twitter tended to ``engage earlier and correct more'' for crisis rumor debunking \cite{starbird2018engage}. Similarly, a study of Weibo found that influencers were more likely to post tweets that debunked COVID-19-related conspiracy theories than ordinary users \cite{Chen2020ConspiracyAD}.
In this paper, we systematically compare four different kinds of debunkers, covering citizen media, official media, and third-party fact-checkers. Past studies have shown that Chinese social media users have different levels of trust in the fact-checking information from these sources \cite{lu2020government}. Our work supports this view by uncovering differences in the behavior and effectiveness of these debunkers.

%
%
%

%
%
%

%

%

%
%

%
%

%

%

%

%


%
\section{Data Preparation}
Sina Weibo\footnote{\url{https://weibo.com/}} is one of the largest microblogging websites in China, where, similar to Twitter, users can post their messages, pictures and videos publicly for instant sharing, 
while other users can retweet, like, and comment on these posts. 
Weibo acts as a central hub for Chinese Internet users to access, disseminate, and receive information and news with 560 millions monthly active users in recent years \cite{weibo_report}. %
This section presents how we 1) collected Weibo posts and health rumors related to COVID-19; 2) extracted posts related to health rumors; and 3) distinguished posts with different behaviors (\ie discussion or debunking). %

\subsection{Data Collection}



\begin{table}[t]
\centering
\begin{tabular}{lll}
\hline
 & Verified information                      & Examples    \\ \hline
Indv & \makecell[tl]{Unverified users, \\ verified individuals   }                            & \makecell[tl]{\textit{Wuhan residents}, \\ \textit{celebrities}}       \\
Org  & \makecell[tl]{Enterprises, institutions, \\ apps, schools, websites} &\makecell[tl]{\textit{Dr. DingXiang},\\ \textit{universities}} \\
Gov   & Government agencies & \textit{Police departments}  \\
Media & Media                  & \textit{Xinhua News} \\ \hline
\end{tabular}%
\caption{Categorical schemes and examples for individuals, organizations, governments, and media.}
\label{tab:account_categories}
\end{table}

We use a dataset of COVID-19 related Weibo posts with user information from Jan 1st, 2020 to May 1st, 2020. 
The dataset was provided by Qingbo Big Data Technology Co Ltd\footnote{\url{http://yuqing.gsdata.cn}} through monitoring posts on Weibo and then filtering out unrelated posts using a series of COVID-related keywords, \eg pneumonia and virus.
The resulting dataset contains 100,159,355 unique posts published by 26,927,690 unique users in total. Each post entry contains textual content, a timestamp, whether it is a retweet, whether it has images or a video attached,
and whether the ID of the post and author can be used to indicate uniqueness. %

We further categorized these unique user accounts into four types, namely \textit{individual}, \textit{organization}, \textit{government}, and \textit{media}, based on account verification information on Weibo. The platform assigns and displays a categorical code to every account to denote their revealed identities, visible to all Weibo users. %
We merged the categories into four types based on user status and occupation, as shown (Table \ref{tab:account_categories}). %

\subsection{Identifying Health Rumors and Related Posts}\label{method:ident_rumors}

\begin{table}[t]
\centering
\begin{tabular}{lcccccc}
\hline
Type      & Range   & Mean     & Sd  & Total     \\ \hline
Wish      & 1 - 38247 & 1147.91     & 3744.45 &312,232 \\
Dread    & 1 - 51941 & 2291.21  & 6587.82 & 311,605\\ \hline
\end{tabular}%
\caption{
Number of rumor-related posts per set in the wish (N=272) and dread (N=136) categories, respectively.
}
\label{tab:set-stats}
\end{table}

To further extract posts related to health rumors from the initial dataset, we first compiled a comprehensive list of pandemic-related rumors circulating in Chinese social media from the archives of eleven popular Chinese fact-checking websites. 
Ten of them are located in mainland China (six are run by commercial companies and four by governments), and one is based in Taiwan. We scrapped 5,958 fact-checking articles published from Jan 2 to April 8, 2020
from these websites. 
We manually inspected the topic and veracity of each article and removed any that were confirmed as not being rumors by these sites.
Next, two annotators (P1 and P2) -- native Chinese speakers and familiar with our work -- identified health rumors from the list (Cohen's kappa $\kappa$ = 0.86) and labelled their types (\ie \textit{wish} or \textit{dread}) according to the definitions in \cite{difonzo2012rumors}  independently ($\kappa$ = 0.81). 
The annotators resolved the conflicts via negotiation. %

\begin{figure}[t]
\includegraphics[width=\linewidth]{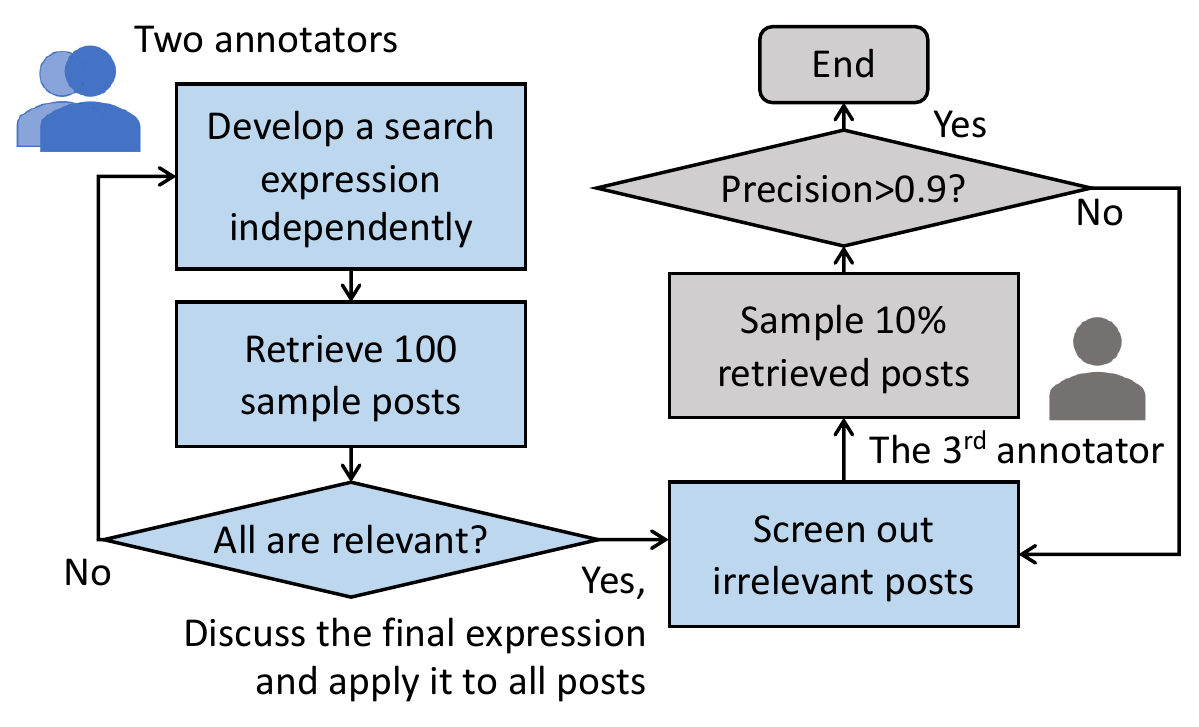}
\centering
\caption{The workflow of identifying all posts related to a rumor topic from the Weibo dataset.}
\label{fig:annotationProcess}
\end{figure}

Following the approach of \cite{shin2017political}, we use logical expressions (\eg \textit{Banana AND Prevent AND (Corona OR Virus)}) to retrieve rumor-related posts via Elasticsearch\footnote{\url{https://github.com/elastic/elasticsearch}}. 
Figure \ref{fig:annotationProcess} illustrates the retrieval process.
For each rumor topic, P1 and P2 came up with initial expressions and manually inspected 100 random samples of the retrieved results to verify their effectiveness. They compared the expressions in the final version and further screened the results returned by these queries as follows.
For Weibo posts extracted in correspondence to each of the health rumors, P1 and P2 separately scanned their contents and eliminated the irrelevant ones.
A third annotator (P3) sampled 10\%
If the precision failed to reach 0.9 
, P1 and P2 needed to repeat the screening step. We did not use recall as a quality control criterion as it was guaranteed as much as possible by the logical expression adjustment step. 
After dozens of rounds of iterative screening,
we ended up with 408 sets of rumor-related posts (623,837 in total), each set corresponding to one health rumor topic. Table \ref{tab:set-stats} shows the descriptive statistics of wish and dread posts. In the following, the term \textit{set} or \textit{post set} denotes a group of posts on the same topic.

\subsection{Categorizing ``Discuss'' and ``Debunk'' Posts}\label{method:categorize_rumor}

\begin{table}[t]
\centering
\begin{tabular}{lcc}
\hline
      & User Count    & Debunking Post Count    \\ \hline
Individual  & 94,127      & 122,448           \\
Government   & 8,675       & 40,225            \\
Media & 2,404       & 8,404             \\
Organization   & 3,502       & 25,534            \\ \hline
\end{tabular}%
\caption{Summary statistics for debunkers and their debunking posts.}\label{tab:debunker-stats}
\end{table}

We divided all rumor-related posts obtained in the previous step into two exclusive classes based on their behaviors: \textit{debunk} (counter-rumor microblog messages) and \textit{discuss} (the rest). 
According to our observation, debunking posts tend to explicitly declare their nature by using denials (\eg \textit{this is not true}) or flagging the misinformation (\eg \textit{this is a rumor}). We thus applied keyword matching to differentiate \textit{debunk} posts from \textit{discuss} posts.
Three annotators first familiarized themselves with randomly selected sets of posts (five each from wish and dread types) and collectively derived an initial list of regular expressions of debunking indicators. %
We then automatically marked all posts with such indicators across all the sets.
After that, we used the same screening procedure mentioned earlier (10\%
and precision threshold 0.9) to iteratively improve the results.
Eventually, we identified a total of 238,554 debunking posts in the post sets, some of which appear in more than one set. These posts were created by 108,708 distinctive debunkers
(\ie the authors of these posts, see summary statistics in Table \ref{tab:debunker-stats}). We treated all remaining posts as discussion posts, which could include affirming or neutral information about the health rumors. %
%

%
%
%

%
%
%
%

%
%
%

%
%
%

%
%

%
%
%
%
%
%
%
%

%

%
%
%
%

%

%

%
%
%
%
%
%
%
%
%
%
%

%
%
%
%
%
%
%

%

%
%
%
%
%
%
%
%
%
%
%

%
%
%
%
%
%
%
%
    
%
%
%
%
%
%
%
%
%
%
%
%
%
%
%
%

%

%
%
    
%
%
%

%
%
%
%
%
%
%
    
%
    
%
%


\section{Characteristics of Health Rumors (RQ1)}

In RQ1, we compare the characteristics of wish and dread types of health rumors in terms of their contents and dissemination patterns.

\subsection{What Concerns were Reflected in Health Rumors?}

\begin{figure}[t]
\includegraphics[width=\linewidth]{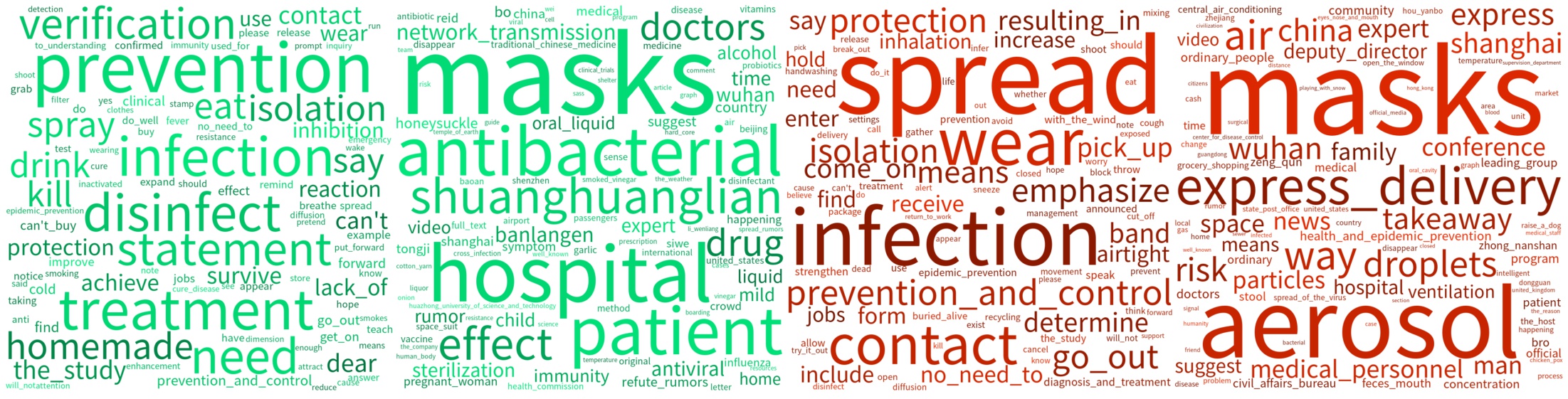}
\centering
\caption{Word clouds for health rumors, from left to right, are verbs and nouns for wish rumors and those for dread rumors; word sizes are determined by the aggregated weights in LDA.}
\label{fig:RQ1-content}
\end{figure}

To understand what kind of public concerns are reflected in the two types of health rumors, we used Latent Dirichlet Allocation (LDA) \cite{blei2003latent} to extract potential topics from rumor-related Weibo posts and visualize them by word clouds. We build different LDA models for each of the two types of rumor-related posts and then select the appropriate number of topics based on the coherence measures \cite{roder2015exploring}. 
A coherence score has no obvious improvement after increasing the number of topics to more than eight for dread rumor-related posts ($\text{score}=0.528$), whereas the most appropriate number of topics as suggested by coherence score is 20 for wish rumor-related posts ($\text{score}=0.510$). 
We observed that the top keywords under the topics identified by LDA are primarily verbs and nouns. Verbs typically indicate the types of activities people became more engrossed in during this public health crisis, and nouns convey the subjects of public attention.
We thus selected the top 10 words from each topic and plotted separate verb and noun word clouds for wish and dread rumors, respectively. To stress the differences between these two types of rumors, we removed high-frequency words (i.e., ``virus'', ``corona'', ``pneumonia'') that are common in both rumor types and then used the aggregated weights of remaining words across their associated topics to determine their font sizes in the word clouds.
\subsubsection{Public Concern over Rumor Topic}
As shown in Figure \ref{fig:RQ1-content}, from the aspect of verbs, the words with the highest weights from wish rumor-related topics included ``prevention'', ``treatment'', and ``disinfect'', which suggests that people are primarily interested in how to prohibit, fight against, and recover from coronavirus when discussing wish rumors.
In contrast, the top words in the dread rumor-related topics include ``spread'', ``infection'', and ``contact''. It reflects the general public's concern about the transmission and infectivity of the virus when talking about dread rumors. 
In terms of nouns in the extracted topics, ``(face) mask'' gained the most attention, both as a source of dread and as a source of wish, but from different perspectives. Wish rumors generally described how to ``get'' masks, such as DIY masks made with homemade materials, while the dread ones focused on the disposal of used masks, \ie warning people to throw away masks used once under any circumstances. Such heated discussions around these themes may have been due to the scarcity of masks in China at the beginning of the public health crisis \cite{scmp_2020}.
The top nouns in wish rumor topics also include ``antibacterial (substance)'', ``huanghuanglian'', ``hospital'', ``doctor'', ``drug'', and ``alcohol'', showing that people may tend to gain hope from these entities. On the contrary, the top nouns contained in the dread rumor topics include ``aerosol'', ``express delivery'', ``air'', and ``droplets'', showing that the media of virus transmission often triggers fear in people.
\subsection{How the Spread of Health Rumors Varied by Type?}

We next compare the spread characteristics of the two kinds of health rumors. 
We first use a variety of single-variable linear regression models to examine the relationship between health rumors and their overall dissemination patterns. The way of using regression analysis to characterize different types of research subjects (e.g., post and user types) has been widely adopted in past work \cite{yang2016did, Chen2020ConspiracyAD}. In our case, the independent variable (IV) of each model is a categorical variable representing the type of rumors with the ``wish'' type as the reference group, i.e., $\text{wish}=0$ and $\text{dread}=1$. The dependent variables (DVs) include the following indicators. 
The number of observations is 408 (\ie the total number of post sets). Following \cite{vosoughi2018spread}, we also use curve charts to visualize the differences across rumor types in term of propagation speed.
\subsubsection{Dependent Variables}
One viral nature of online misinformation is that it can be widely spread by many users over a short period of time. Following previous work \cite{starbird2014rumors,arif2016information}, we measured both the volume of \textbf{posts} and \textbf{users} for each health rumor by counting the post ID and the user ID. We also measured the \textbf{duration} by comparing the number of minutes between the earliest post and the last post in each post set. Since these three DVs are highly skewed, we apply a log transformation on them before analysis.
Moreover, rumors threaten people's mental health by spreading panic on social networks \cite{ahmad2020impact}. We quantify such  \textbf{negativity} by computing the proportion of negative posts\footnote{By sentiment analysis API: \url{http://databus.gsdata.cn/open}} in each post set. 
\subsubsection{Macro Perspective}

\begin{table}[t]
\centering
\fontsize{9}{10}\selectfont 
\begin{tabular}{lllll}
\hline
                     & Post     & User     & Duration   &   Negativity \\ \hline
Dread                & 0.394*** & 0.397*** & 0.358***   &   0.081** \\ \hline
Intercept            & 2.099*** & 2.055*** & 2.054***   &   0.450*** \\
$R^2$ & 0.039    & 0.040    & 0.037 & 0.024    \\ \hline
\end{tabular}%
\caption{Prediction of dissemination patterns of health rumors. Wish type serves as the reference category. ***:$p<0.001$; **:$p<0.01$; *:$p<0.05$. N=408.}
\label{tab:RQ1-static_ch}
\end{table}

Regression analysis results (Table \ref{tab:RQ1-static_ch}) show that,
dread rumors involved significantly more posts ($\beta$ = 0.394, $p$ $<$ 0.001), users ($\beta$ = 0.397, $p$ $<$ 0.001) and lasted longer than wish rumors ($\beta$ = 0.358, $p$ $<$ 0.001). 
Proportions of negative posts of dread rumors were higher than wish type ($\beta$ = 0.081, $p$ $<$ 0.01). 
\subsubsection{Dynamic Perspective}\label{sec:RQ1-dyn}

\begin{figure}[t]
\includegraphics[width=\linewidth]{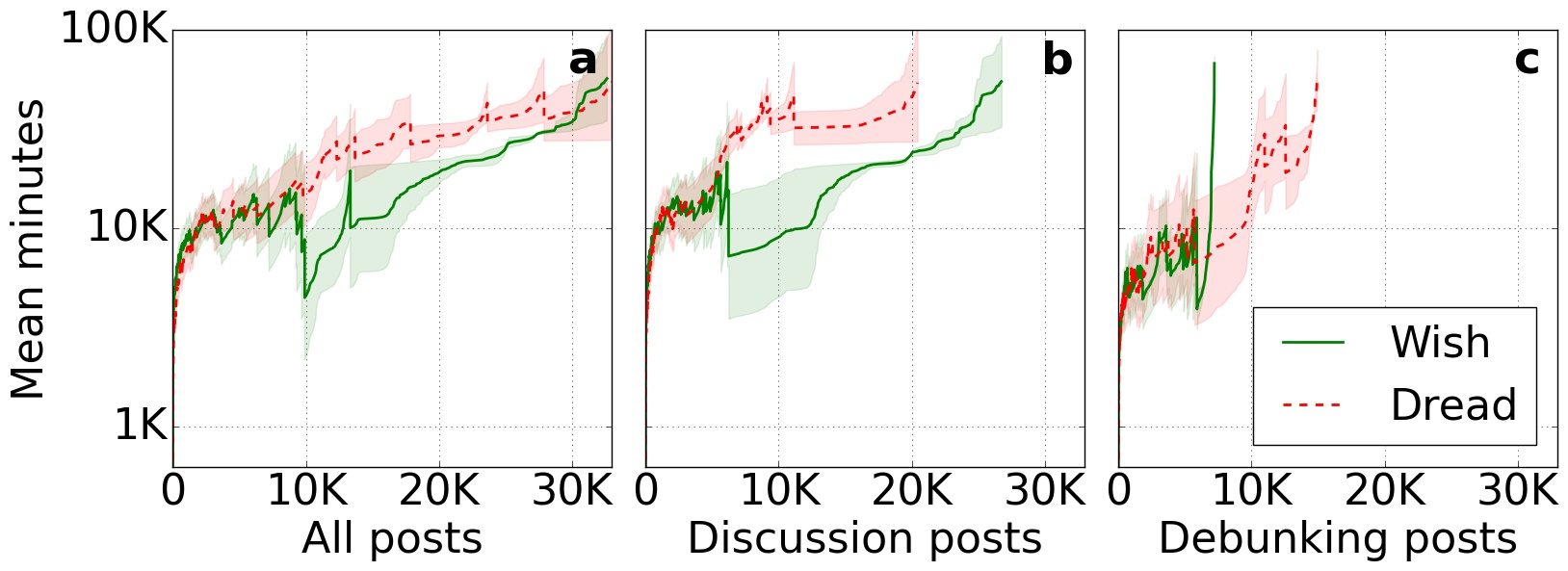}
\centering
\caption{
The average minutes it takes for wish vs dread rumors to reach any number of a) posts b) discussion posts c) debunking posts; ±1 standard errors are drawn as shadow. }
\label{fig:RQ1-dynamicv2}
\end{figure}

Figure \ref{fig:RQ1-dynamicv2} demonstrates the temporal dynamics of the spread of wish and dread rumors. These graphs illustrate the amount of time (y-axis: minutes) needed for a (wish/dread) health rumor to reach a certain level of exposure in Weibo (x-axis: number of related posts). In particular, in Fig \ref{fig:RQ1-dynamicv2}a, one can see that it takes a similar length of time for wish and dread rumors to have up to 10K Weibo posts mentioning them. In fact, 96.5\% of the rumors in our dataset reached fewer than 10K related posts during the four-month period.
For the remaining 3.5\% of extreme rumors (hitting greater than 10K posts in Fig \ref{fig:RQ1-dynamicv2}a), one can clearly observe that these wish rumors spread 
faster than the dread ones. 
The green line (wish) is noticeably lower 
than the red line (dread), means that on average less time is spent to reach the same amount of exposure.
We further split the posts based on their nature into two subsets -- discussion and debunking, and measured their speed of propagation separately. %
Overall, the growing trend of discussion posts (Fig \ref{fig:RQ1-dynamicv2}b) is similar to that in \ref{fig:RQ1-dynamicv2}a with an even wider gap between dread and wish rumors.
However, in Fig \ref{fig:RQ1-dynamicv2}c, we found that the emergence of wish-related debunking posts slows down drastically after the post number reaches around 6.8K and nearly stops at 7K. 
This phenomenon comes from two ``extreme'' wish rumors (``Shuanghuanglian'' and ``alcohol can cure/prevent the coronavirus''). They are the wish rumors that received the most debunking responses in our dataset. 
At the beginning, their debunking posts reached around 6.8K within a week, then rapidly dropped to less than ten posts a day in the following two months. 
Instead, debunking posts related to dread rumors, such as ``takeaway food spreads the coronavirus'', continued to appear in around dozens to a hundred posts per day.
This indicates that while dread rumors have been continuously debunked by the public, efforts on countering wish rumors died down after some time.
This might lead to wish rumors later being disseminated faster than dread rumors. We further analyze the specific effect of debunking on suppressing both types of rumors quantitatively in reply to RQ3. 
To sum up, by identifying characteristics of wish and dread rumors during the pandemic (RQ1), we found differences exist between these two types of rumors in terms of their content and dynamics, which inform the rumor management agents to treat them separately according to their distinct patterns.
%
%

%
%

\section{Identification of Debunking Efforts (RQ2)}

RQ2 examines how different groups of social media users engage in dispelling health rumors during COVID-19 and the strategies they prefer to use.

\subsection{How do Debunkers Engage Differently in Counter-rumor Activities on Weibo?} %

We built a series of single-variable linear regression models to investigate the relationship between the type of debunkers and their engagement. Model IV consists of four categories representing different debunker types, with the ``individual'' type as the reference. We use the following indicators as DVs. We further use descriptive statistics, such as estimated mean $\mu$, to quantify the degree of variations. 
We also apply the t-test and Kolmogorov-Smirnov (KS) test to evaluate if the differences in terms of means and distributions are statistically significant.
The number of observations is 108,708 (\ie the total number of debunkers).

\subsubsection{Dependent Variables}
Debunking volume and timing are essential aspects of previous studies on user debunking behavior \cite{starbird2018engage}. Likewise, we analyze the \textbf{volume} and average \textbf{timeliness} of debunking posts made by the user over the four-month period. The timeliness of a debunking post is measured by the amount of time (in minutes) between it and the first post in that post set.

\begin{table}[t]
\centering
\fontsize{9}{10}\selectfont 
\begin{tabular}{lll}
\hline
                     & Volume & Timeliness \\ \hline
Government                  & 0.245***        & \textbf{0.202}***           \\
Media                & \textbf{0.480}***        & 0.177***           \\
Organization         & 0.104***        & 0.108***           \\\hline
Intercept            & 0.335***        & 3.960***           \\
$R^2$ & 0.258           & 0.015              \\ \hline
\end{tabular}%

\caption{
Prediction of the engagement of debunkers. Individual serves as the reference category. ***:$p<0.001$; **:$p<0.01$; *:$p<0.05$. N=108,708. %
}
\label{tab:RQ2-RA-engage}
\end{table}

\begin{table*}[t]
\centering
\fontsize{9}{10}\selectfont 
\begin{tabular}{llllll|llll}
\hline
               & \multicolumn{5}{c|}{Functional Preferences}                                                      & \multicolumn{4}{c}{Rhetorical   Preferences} \\
               & Retweet   & Duplicate & Image             & Video     & Length\footnotemark & Fact      & Emotion   & Sarcasm   & Source   \\ \hline
Government &
  -1.074*** &
  \textbf{0.743}*** &
  0.547*** &
  \textbf{0.905}*** &
  0.095*** &
  \textbf{3.205}*** &
  -1.920*** &
  \textbf{-3.628}*** &
  1.092*** \\
Media &
  \textbf{-3.452}*** &
  0.419*** &
  0.718*** &
  0.708*** &
  \textbf{0.253}*** &
  2.687*** &
  \textbf{-2.653}*** &
  -2.410*** &
  \textbf{1.899}*** \\
Organization   & -1.954*** & -0.038    & \textbf{1.075}*** & 0.254***  & 0.066***                                 & 2.547***  & -2.511*** & -2.976**  & 1.079*** \\ \hline
Intercept      & 0.043***  & 0.455***  & -1.173***         & -1.031*** & 2.128***                                 & 1.050***  & -0.667*** & -1.328*** & 0.074*** \\
(Pseudo) $R^2$ & 0.121     & 0.015     & 0.02              & 0.027     & 0.072                                    & 0.18      & 0.171     & 0.164     & 0.104    \\ \hline
\end{tabular}
\caption{
Prediction of debunkers' functional preferences (N=238,554) and rhetorical preferences (N=1,025). Individual serves as the reference category. ***:$p<0.001$; **:$p<0.01$; *:$p<0.05$.
}
\label{tab:RQ2-RA-preference}
\end{table*}

\subsubsection{Engagement}
Results of regression models are presented in Table \ref{tab:RQ2-RA-engage}.
Compared to individuals, government ($\beta$ = 0.245, p $<$ 0.001), media ($\beta$ = 0.480, p $<$ 0.001), and organization ($\beta$ = 0.104, p $<$ 0.001) all posted significantly more debunking posts in Weibo during the pandemic. 
The results of KS-test 
and t-test show their differences are significantly in the shape of distributions and mean (both p $<$ 0.001), and all debunkers participated more in correcting dread rumors: Individuals ($\mu_\text{wish}$ = 0.55, $\mu_\text{dread}$ = 0.73), Government ($\mu_\text{wish}$ = 2.01, $\mu_\text{dread}$ = 2.544), Media ($\mu_\text{wish}$ = 4.739, $\mu_\text{dread}$ = 5.719), and Organization ($\mu_\text{wish}$ = 1.134, $\mu_\text{dread}$ = 1.226).
Regression results of timeliness in Table \ref{tab:RQ2-RA-engage}  show that individuals are the fastest users to publish debunking posts 
due to the other three having a relatively positive $\beta$. 
We also calculated the $\mu$ of the debunking of posts' timeliness (converted to days) from these debunkers: Individual (10.11 days), Media (12.90 days), Organization (12.73 days), and Government (13.26 days). It shows that individuals are more likely to engage in debunking activities around three days earlier than other debunkers. We further compared the timeliness of debunkers engaging in wish and dread rumors. KS test and t-test reveal that individuals and government show significant differences in timeliness between wish and dread rumors (both p$<$0.001). On average, individuals and governments were later to debunk dread rumors than wish rumors. Their estimated mean (convert to days) are: Individuals ($\mu_\text{wish}$ = 9.069, $\mu_\text{dread}$ = 10.966) and Government ($\mu_\text{wish}$ = 12.126, $\mu_\text{dread}$ = 14.075). 

\footnotetext{Using linear regression}

\begin{table*}[t]
\centering
\begin{tabular}{llllllll}
\hline
 &            & \multicolumn{2}{c}{Wish}                    & \multicolumn{2}{c}{Dread}                     & \multirow{2}{*}{KS} & \multirow{2}{*}{t} \\
 & Preference & \multicolumn{1}{c}{$\mu$} & \multicolumn{1}{c}{SE} & \multicolumn{1}{c}{$\mu$} & \multicolumn{1}{c}{SE} &                     &                    \\ \hline
Government & Retweet & \textbf{0.292} & 0.003 & 0.248          & 0.003 & 0.022*** & 4.891***   \\
Individual & Image   & \textbf{0.359} & 0.002 & 0.200          & 0.001 & 0.074*** & 30.091***  \\
Media      & Image   & \textbf{0.445} & 0.004 & 0.364          & 0.004 & 0.050*** & 8.068***   \\
Individual & Video   & 0.176          & 0.001 & \textbf{0.310} & 0.002 & 0.087*** & -34.175*** \\
Individual & Length  & \textbf{186.0} & 1.035 & 166.2          & 0.760 & 0.087*** & 27.439***  \\
Government & Length  & \textbf{292.7} & 3.203 & 245.7          & 2.380 & 0.098*** & 20.996***  \\
Media      & Length  & \textbf{413.9} & 4.990 & 314.1          & 3.174 & 0.118*** & 21.140***  \\ \hline
Individual & Satire  & \textbf{0.264} & 0.025 & 0.107          & 0.025 & 0.157*   & 3.998***   \\
Individual & Source  & 0.472          & 0.029 & \textbf{0.610} & 0.039 & 0.138*   & -2.841**   \\ \hline
\end{tabular}
\caption{Statistics and hypothesis tests for function preferences (the top part) and rhetorical preferences (the bottom part) of debunking posts created by different types of debunkers relating to wish versus dread rumors. ***:$p<0.001$; **:$p<0.01$; *:$p<0.05$.}
\label{tab:RQ2-test-Pref}
\end{table*}

\subsection{How did Posters' Preferences of Debunking Strategies Differ?}
We then examine each debunking post created by users to investigate the association between their types and their preferences when editing posts, again using single-variable regression analysis with the user type as IV.

\subsubsection{Dependent Variables}
The meta data of the debunking posts are used to indicate users' \textit{functional preferences}, such as whether it is a \textbf{retweet}, whether it includes an \textbf{image} or \textbf{video}, and the \textbf{length} of the text. 
Users can also republish a post in a different way than through Weibo's retweet feature by copying and pasting it. Users may have different motives for such behavior, such as to have their posts appear as ``original''. We capture such a preference by SimHash\footnote{\url{https://github.com/yanyiwu/simhash} The same hash value means having the same (or only subtle modifications) textual content with others.} to see if a non-retweeted post is a duplicate in content to an earlier post.

For a more fine-grained description of debunking preferences,
we adapt Aristotle’s modes of persuasion (\ie \b{Logos}, \b{Pathos}, \b{Ethos}) to describe how debunking messages are constructed using \textit{rhetorical strategies} to change the attitude or behavior of others toward rumors \cite{petty2012communication}. %
The two researchers first constructed the codebook of rhetorical strategies by a mix of inductive and deductive coding of 200 randomly sampled debunking posts. Another 1,025 posts were sampled for formal coding based on the codebook. Each post could be coded by multiple strategies.
The percentage of coded strategies among the 1,025 debunking posts were: stating \textbf{facts} (88\%), expressing strong \textbf{emotions} (18\%), \textbf{sarcasm} (10\%), and citing credible \textbf{sources} (70\%); corresponding Cohen’s kappa scores are 0.509, 0.699, 0.692, and 0.738.
Four dichotomous variables are used to represent these strategies.

As the values of these DVs are either zero or one, we elect to use logistic regression rather than linear regression in this case. However, length is a continuous variable and we retain linear regression for it. Functional preferences are analyzed using 238,554 observations (all debunking posts), whereas rhetorical preferences are analyzed using 1025 observations (sampled debunking posts).

\subsubsection{Functional Preferences} \label{sec:RQ2-func}

The regression results are shown in Table \ref{tab:RQ2-RA-preference}. 
Compared to individuals, the other three types are significantly less likely to use the retweeting function ($\beta$ $<$ 0, $p$ $<$ 0.001).
In addition, compared to individuals, media ($\beta$ = 0.419, $p$ $<$ 0.001) and government ($\beta$ = 0.743, $p$ $<$ 0.001) were significantly more likely to publish text content duplicated from some previous posts. %
By examining samples of duplicate posts from government debunkers, we found that they tend to quote posts from media, especially from high-impact media involving a government background, \eg \textit{@People's Daily}. Their quoted content can be in diverse formats, including texts, images, and videos, and are usually tagged with their original sources, for example:
\textit{``[Will shoes bring the virus home?]...no need to disinfect the soles of shoes in daily life...@People's Daily.''} 
Combined with our previous finding that governments tended to engage in debunking activities at a later time, this may indicate the cautious attitude of governments when handling health rumors; they 
may like to
wait until things become clearer
before quoting and posting information. 
Compared to individuals, the other three types were more likely to use multimedia (i.e., images and video) in their debunking posts ($\beta$ $>$ 0 with individual as the reference in regression analysis for both DVs of image and video, all of $p$ $<$ 0.001). 
According to our observation, 
many debunkers chose to put arguments and explanations in pictures to avoid the character limit\footnote{Unlike Twitter, Weibo allows posts longer than 140 characters but requires an additional click to expand the full text.} of Weibo. They also preferred to put interviews with medical experts in the attached videos, and encourage audiences to watch the videos for more details. For example, \textit{``Can bathing with hot water at 56°C fight the virus?.... Click on the video ↓ Experts have all the answers for you!''} %
Compared to individuals, all the other users preferred to write longer content in their counter-rumor posts ($\beta >$ 0, $p<$0.001).

Between wish and dread rumors,
individuals were more likely to send videos in dread-related debunking ($\mu_\text{dread}$ = 0.31, $\mu_\text{wish}$ = 0.176, $p$ $<$ 0.001 for both KS-test and t-test), and all debunkers except organizations ($p$ $>$ 0.05 for KS-test) tended to write longer texts when fighting wish rumors than they did with the dread ones ($\mu_\text{wish}$ $>$ $\mu_\text{dread}$, $p$ $<$ 0.001 for both tests). We present the detailed results in Table \ref{tab:RQ2-test-Pref}. %

\subsubsection{Rhetorical Preferences} 
%

%

%
%
%
%
%

%
%
%
%
%
%
%
%
%
%
%
%
%
%
%
%
%
%
%


%
%
%
%
Table \ref{tab:RQ2-RA-preference} also summarizes the results of regression analysis on 1,025 sampled debunking posts.
Overall, the results show that government and media were significantly more likely to use facts, such as evidence and reasoning, in their debunking information compared to individuals ($\beta$ $>$ 0, $p$ $<$ 0.005).
For example,
\textit{``[Can antibiotics be used to treat coronavirus?]...No, antibiotics are ineffective against viruses, only bacteria. 2019-CoV is a virus and therefore antibiotics should not be used as a means of prevention or treatment.''
} 
%
%


%
%
Compared to individuals, the other three kinds of debunkers were significantly less likely to use strong emotion and sarcasm in their health rumor debunking posts ($\beta$ $<$ 0, $p$ $<$ 0.005). An example message from an individual debunker is:
\textit{``That shuanghuanglian is only suppress! Suppress! You have to be sick to be useful...If you get sick, the hospital will give it to you, for free!''} 
In contrast to individuals, all the other debunkers had a significantly higher tendency to cite credible sources ($\beta$ $>$ 0, $p$ $<$ 0.005) in their posts. 
We note that one typical counter-rumor strategy adopted by media was showing interviews or quotes from medical experts. The media usually specified the expert's name, affiliation, and qualification in the text. For example:
\textit{```No evidence indicates that coronaviruses will disappear in summer', Michael Ryan, Executive Director of the WHO's Health Emergency Programme, said at a regular press conference in Geneva.'' 
}
This could reflect the rigorous attitude of the media toward the sources of information.

Between wish and dread rumors,
we only found significant differences in the use of sarcasm and credible source by individual debunkers in countering these two types of rumors ($p$ $<$ 0.05 for both KS-test and t-test). According to results in Table \ref{tab:RQ2-test-Pref}, individuals were significantly more likely to apply sarcasm on debunking wish rumors than dread rumors ($\mu_\text{wish}$ = 0.264,  $\mu_\text{dread}$ = 0.107). For instance, 
\textit{``Rumors are everywhere.
Some say that the virus cannot live if you set air conditioning at 20°C, ..., I can also think of one, eating spicy chips can prevent viral infection.''}
This suggests that individuals may treat these wish rumors, but not so much the dread ones, in a joking manner.
Besides, individuals referenced reliable sources to support their point when fighting dread rumors significantly more than when they were faced with wish rumors ($\mu_\text{wish}$ = 0.472, $\mu_\text{dread}$ = 0.61). For instance: 
\textit{``WHO says there is no evidence that dogs and cats can get and spread COVID-19. People who abandon their pets, can't you read?!!!''}

To sum up, by comparing the debunking efforts of four kinds of social media users (RQ2), we found that individual debunkers seemed to engage earlier in health rumor combating and retweet the most; organizations preferred to use images to debunk rumors; media tend to contribute more and longer content; while government accounts were more likely to echo previous posts and use videos. Individuals were more likely to use emotion-based strategies such as strong emotion and satire, while other debunkers tended to use facts and cite sources. This sheds light on the use of crowd wisdom for misinformation control in social media.
%

%
%
%
%
%
%
%
%


\section{Effectiveness of Debunking Efforts (RQ3)}
In RQ3, we investigate the effect of debunking on health-related rumors through a combination of two analytical methods.  
First, we use \textit{Granger causality analysis} \cite{granger1969investigating} to test whether the trends of debunking posts help predict the trends of discussion posts. Second, we introduce a new measure, called \textit{suppression ratio}, to quantify
how declines of rumor discussion posts vary with different factors -- rumor types and user roles in particular -- and over time.

\subsubsection{Granger Causality Analysis}

Granger causality analysis \cite{granger1969investigating} is a statistical hypothesis test that can be used to determine the relationships between two variables by checking whether or not a time series of variable $X$ (in this case, the trend of debunking posts) is useful in predicting variable $Y$ (the trend of discussion posts). This test is accomplished by using $n$ number of lags (\ie previous values) of $X$ to model the change in $Y$. A $p$-value from the test is used to determine whether the null hypothesis that $X$ does not help predict (\ie Granger-cause) $Y$ should be rejected. 
Since our data are grouped into different rumor topics, we tested the Granger causality between debunking and discussion on each topic separately.
Nevertheless, to demonstrate the impact of rumor debunking across different topics, we follow \cite{dutta2018measuring} to report the proportion of rumors with statistically significant results (\ie $p<0.05$) out of the entire rumor pool. %

The specific steps for testing Granger causality under each rumor topic are as follows. We first obtain a pair of time series $X$ and $Y$ from social media data related to a given health rumor topic by counting the number of debunking posts and discussion posts within every six hours. 
Since rumor propagation usually occurs within a short period, selecting a long sampling period, for instance, one day, may lead to an aggregated time series that is too short and obscures the short-term information.
Similar to \cite{de2017computational}, 
we remove topics whose series contain a gap of more than one week between two consecutive steps.
Following \cite{lutkepohl2005new},
we convert the time series into a stationary series and remove the rumor topics if the resulting series fail to satisfy the Augmented Dickey-Fuller (ADF) test ($p<0.05$) or are found to be autocorrelated by Durbin-Watson Statistic.
With the remaining data, for each pair of time series that corresponds to one rumor topic, we construct a vector auto-regressive (VAR) model to determine the optimal lag $n$ required by the Granger causality test.
The VAR model is used to predict the $Y$ series using $n$ lags of the $Y$ series and $n$ lags of the $X$ series. 
We select the lag value $n$ that maximizes the likelihood of the VAR model to make the most accurate prediction as in \cite{de2017computational}. In our case, the best lag value indicates that the past $n$ steps of the debunking time series contain the most useful information for predicting the discussion series. We report the distribution of these best lags based on the VAR model across rumors in Figure \ref{fig:RQ3-lags}, which in a sense reflects the time span of the debunking effects.

\subsubsection{Results of Granger Causality Analysis}

\begin{figure}[t]
\includegraphics[width=\linewidth]{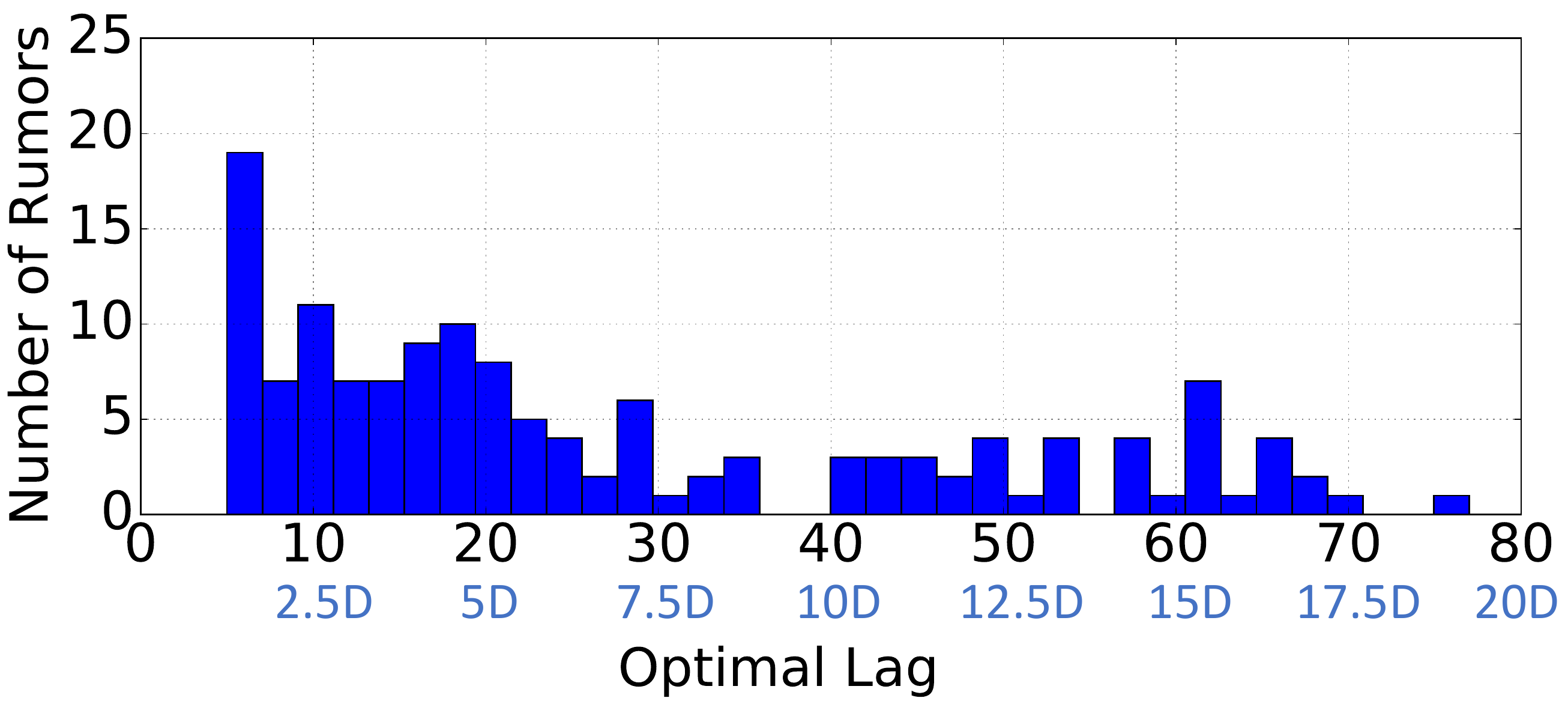}
\centering
\caption{The histogram of the optimal lags derived from using VAR models to predict the trend of rumor discussion posts based on the trend of debunking posts.
Each lag represents six hours of information. The lag value is also provided in days (D).}
\label{fig:RQ3-lags}
\end{figure}

In total, 144 (90 pairs from wish rumors and 54 pairs from dread rumors)
were obtained and analyzed using the Granger causality test.
We observed significant ($p<0.05$) Granger causation between debunking and rumor discussion in 111 health rumors (77.1\%).
Separately, such causation is observed in 69 wish-type rumors (76.7\%) and in 42 dread-type rumors (77.8\%).
This result suggests that debunking and rumoring is closely related, and debunking changes are useful in predicting the changes in rumor discussion. 
Figure \ref{fig:RQ3-lags} shows the distribution of the optimal lags from 144 constructed VAR models. 
In about 10\% of health rumors, the best lag is 7, meaning that combining the past seven steps (\ie $6\times7$ hours or 1.5 days of past information) of the debunking and discussion series can predict the subsequent discussion series the best. This is compared to combining other lengths of past information. Moreover, 60\% of the best lags are $\le$ 24 (6 days), and 75\% of the optimal lags are $\le$ 30 (7.5 days). These results suggest that using information from a relatively short previous time window (usually less than a week) for the debunking series is the most helpful in predicting changes in succeeding discussion series, possibly revealing that the 
association between debunking and rumor dissemination is usually short-term.

To date, our results suggest that debunking can have a short-term impact on the volume of discussions around health rumors. However, the analysis methods we have used do not fully answer the question as to whether debunking to have a suppression effect on the spread of rumors, and whether these effects vary across rumor types and debunker types. We aim to answer these questions through further suppression ratio analysis in the following subsections.

\subsubsection{Suppression Ratio Analysis}

Suppression ratio is a new measure we propose to examine how effectively debunking suppresses the discussion of a rumor.
We first define a post to be ``\textit{effective}'' if the volume of discussion posts (\ie non-debunking posts) on the same topic in $T$ (a given interval) after its publication is no more than that in $T$ before it appears on social media. %
This is consistent with previous ideas of measuring the decline in rumor-related posts after the appearance of debunking posts \cite{andrews2016keeping, shin2017political}. Given that public attention to rumors may naturally decline over time, we further calculate the ratio of effective discussion posts to all discussion posts as the baseline suppression ratio (to reflect the natural decline trend). Similarly, we calculate the suppression ratio of debunked posts. If this ratio is greater than the baseline, we consider 
debunking has a ``\textit{suppression effect}'' on rumor-related discussion. 
By varying $T$ and the source of debunked posts, we can measure how debunking effectiveness is affected by these factors. 

\begin{figure}[t]
\includegraphics[width=\linewidth]{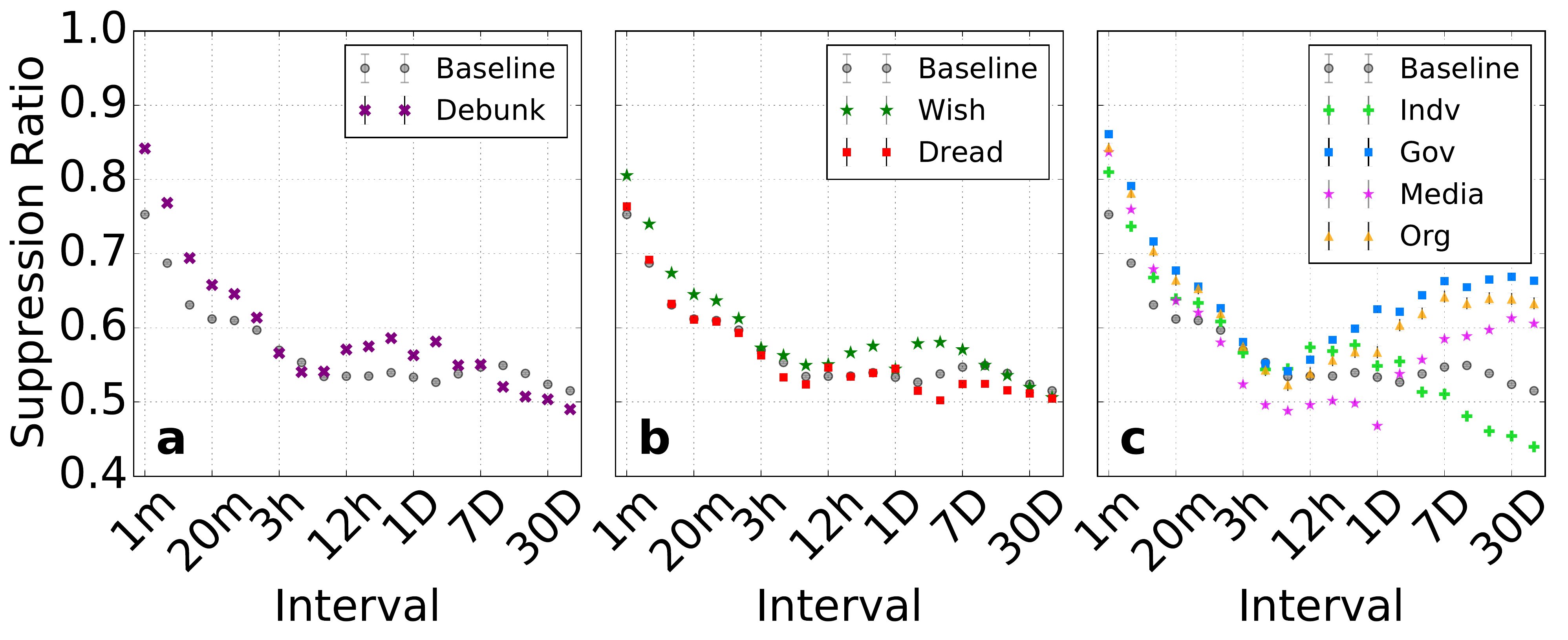}
\centering
\caption{The suppression ratios 
for a) debunking posts, b) two types of health rumors, c) four kinds of debunkers, at different intervals. 
Error bars represent the 95\% CI. 
The units of interval are: minutes (m), hours (h), days (D).
}
\label{fig:RQ3-supression}
\end{figure}

\subsubsection{Suppression Effect}
In Figure \ref{fig:RQ3-supression}, we define ``suppression ratio'' (y-axis) as the percentage of ``effective'' (debunking) posts and the interval (x-axis) denotes a specified length of time before and after posting for assessing suppression effectiveness. In all three subfigures, we use the suppression ratio of discussion (i.e., non-debunking) posts as the baseline, illustrated as grey dots. In general, given an interval \eg 10 minutes, if the suppression ratio of debunking posts is above that of the baseline, we may postulate that it helps reduce the amount of discussion about health rumors to some extent. 

We first investigated the effectiveness of all debunking posts as a whole in lowering health rumor discussions on different time scales. Overall, Fig \ref{fig:RQ3-supression}a shows that the percentage of effective ones among all counter-rumor posts is higher than that of all non-debunking posts in the short-term  ($\le$ 1h) or medium-term (12 hours to 4 days) assessment time window. However, if we compared the amount of discussions before and after the publication of a post on the same topic in a 7+ day window, one can see that the percentage of seemingly effective debunking posts drops below that of non-debunking posts. This may indicate a tendency for a resurgence of rumor discussions for about one week after the debunking activities.

Then, we compared the effectiveness of debunking efforts on wish versus dread type of health rumors. As shown in Fig \ref{fig:RQ3-supression}b, the consistently higher percentage of effort fighting wish rumors seems to result in a decrease in relevant rumor discussions than the baseline rumor-related, non-debunking posts regardless of the length of the assessment period. This may imply a positive influence of debunking activities that target wish rumors on pruning subsequent discussions. Such an observation is consistent with our finding in RQ1 %
that once wish rumor debunking posts cease to occur, the amount of discussions about these rumors grows at a fast rate. In contrast, the ratio of effective posts for combating dread rumors stays the same as or is lower than the baseline. In certain assessment time windows (\eg 1h-10h, 3D-45D), a higher-than-baseline proportion (even $>$ 50\% in a 3+ day window) of dread debunking posts seems to be counterproductive. 

Finally, we compared rumor-suppression effectiveness across different types of rumor debunkers. As illustrated in Fig \ref{fig:RQ3-supression}c, government and organization types of debunkers seem to publish a higher percentage of posts that appear to have a drop in rumor discussions within any interval length after publishing than the baseline. Their proportion of effective debunking posts first decreases as the evaluation window expands from 1 minute to 10 hours, and grows back again as the window size continues to enlarge -- widening their lead over the baseline and the other two types of debunkers. Data of media debunking activities share a similar "U" shape, but with a sharper decline and then a slower come-back as the time frame increases in size. By contrast, the effect of debunking efforts by individuals seems to deteriorate when the assessment window is greater than 18 hours. 
In more than half of the cases, more discussions on the corresponding rumor circulate around Weibo at an interval of one week or longer after the publication of an individual-authored counter-rumor post.

In sum, we found that debunking had a short-term effect on health rumor discussions and inhibited the growth of such discussions.  Debunking efforts by governments and organizations were the most persistent and effective.

%
%
%
%

%
%
%
%

%
%
%
%
%
%
%
%
%
%
%
%
%

%
%
%
%
%
%

%

%
%
%
%

%

%

%


\section{Discussion}
This paper examines how people discuss health rumors and engage in debunking activities on a Chinese social media during the COVID-19 health crisis. Our key findings suggest that: 1) dread rumors generally spread more virally and receive more attention than wish rumors (except for the most extreme ones); 2) when fighting rumors, individual citizens tend to be more emotional while institutions are more critical; and 3) debunking efforts are effective overall, but may sometimes have the opposite effect (\eg counter-rumor posts by media may increase rumor discussions in the short term).

\subsection{Implications}
\subsubsection{Theoretical implication for crisis informatics}
This study provides preliminary large-scale empirical evidence on how various kinds of online health rumors spread and are corrected in the context of health crises. 
First, our findings generally support the fact that dread rumors outnumber wish rumors \cite{difonzo2007rumor} and social media users pay more attention to them than to wish rumors \cite{difonzo2008watercooler, chua2018intentions} under the pandemic setting. The reasons behind the differences are often explained by the negativity bias theory \cite{cacioppo1994relationship}, which suggests that people tend to attach more importance to negative news than to good news. However, our other findings on extreme rumors complicate these views -- in some cases, rumors of hope can attract far more public attention. Future research is needed to investigate whether this is caused by the particular nature of a crisis.
Second, %
we confirm the active role of specific kinds of users in rumor correction, such as journalists \cite{andrews2016keeping}, mainstream media, and official accounts \cite{starbird2018engage}. Our analysis on  rhetorical strategies proposed in persuasion theory \cite{petty2012communication} provides interesting insights into the different debunking mechanisms applied by various social media user groups, \eg the tendency of individuals to use sarcasm.
Finally, as previous works have indicated, the effectiveness of rumor correction is time-sensitive \cite{lewandowsky2012misinformation, ozturk2015combating}. The introduction of a temporal scale of debunking effects could provide a more nuanced perspective on how rumors are suppressed by assorted debunking efforts.

\subsubsection{Practical implication for risk management}
Our findings generally support the suppression effect of debunking activities on the spread of health rumors but also reveal some challenges faced in the process.
On the one hand, health information seekers use social media to gain knowledge (\eg treatments and consequence \cite{gui2017understanding}) and make health decisions \cite{zhang2017knowledge}. However, the high degree of uncertainty associated with a health crise \cite{gui2017managing} and the lay public's lack of expertise make it hard for them to evaluate the quality and veracity of online health information on their own \cite{ortutay2020virus}. In such cases, tips or criticisms from credible platforms and knowledgeable users are valuable signals of information accuracy.
On the other hand, choosing how to debunk rumors can be costly for risk managers \cite{miller2020researchers}. Under unknown situations, the boundaries of information veracity are blurred, \eg whether a homemade face-mask is helpful may depend on the materials. Simply criticising all such information in the face of extreme scarcity of masks may increase public anxiety. Yet, the lack of debunking can in turn, as found in our data, result in the continued spread of rumors. 

For such a dilemma, more care is needed about when and how to perform risk communication in a pandemic. First, as implied in our findings, health authorities and official agencies should establish a good understanding of health-related concerns of the general public from past crisis experiences to address their information needs in a timely manner \cite{gui2017understanding}.
Second, while collective debunking activities on social media are seemingly effective, some users' social network structures are rather closed and exhibit homophily \cite{mcpherson2001birds}. Thus useful information from debunkers may never get circulated to them. Social media platforms may consider strategies to break rumor echo chambers, for example, by automatically attaching counter-rumor posts to previous rumor discussions \cite{ozturk2015combating}.
Third, %
even though people have a tendency to focus on ``bad news'' in a public health crisis, one should not overlook the harm of some inaccurate or false hope. It is important to maintain adequate debunking efforts for both kinds of misinformation. Besides, some risky ways of debunking should be avoided, such as ``post first, check later'' \cite{bruno2011tweet}, because they may have the opposite effect and trigger more discussions about the target rumor, as shown in our data. %

\subsubsection{Generalization}

Some of our findings can be applied to other crisis situations as well as to other social media platforms. In terms of different popularity and spread characteristics across rumor types, dread rumors were also observed more frequently in other health crises such as the Ebola outbreaks \cite{allgaier2015communication, sell2020misinformation}. Similarly, misinformation that provokes fearful emotions have also been found to spread more widely across various rumor topics, \eg politics, financial information, and natural disasters \cite{vosoughi2018spread}, compared to those evoking positive emotions such as trust and joy. Additionally, the active engagement of citizens and government agencies in rumor correction activities is also reported in other social media platforms such as Twitter \cite{andrews2016keeping, starbird2018engage}. However, our findings on debunking effectiveness may be influenced by certain factors specific to the Chinese Internet context, such as people's tendency to trust the government and public institutions \cite{lu2020government}, and thus such results may not be generalized to other global social media. In spite of that, our proposed suppression ratio method for assessing debunking effects is based on the digital trajectories of rumoring and debunking campaigns and can be readily adapted to other data-driven research 
on crisis informatics.

\subsection{Limitations and Future Work}
This paper has several limitations. First, we used keyword-matching and regular expressions to identify COVID-19 related health rumor discussions and debunking posts, which may inevitably miss capturing some data due to complex, noisy usage of language on social media \cite{shin2017political}. %
Second, consistent with similar work \cite{liao2013she}, we classify user accounts into four roles based on the verification information on Weibo. We did not consider more fine-grained user categories. As a preliminary study, we does not distinguish the attitude (pro-rumor or neutral) in rumor discussion posts either. 
Third, the analyses in this study are retrospective and correlational, and therefore cannot determine causation.
Fourth, some of our regression models have low R-squared values, which can be related to the complexity of our research context. Crisis communication in the early pandemic was complicated by many factors, such as information uncertainty and low transparency on Chinese social media. This made it hard to predict rumor propagation and people's debunking behaviors. To reach reliable conclusions, we performed KS- and t-tests and found consistent results of different associations across rumor/user categories.
Fifth, we measure the effectiveness of debunking based on the decline in the number of discussion posts. It is possible that these changes are caused by other factors such as the outbreak of new events diverting public attention \cite{andrews2016keeping, starbird2018engage}. However, we assume that our data volume is large enough to show the dominant relationship between the appearance of debunking and the decline in discussion without being susceptible to other subtle factors. In the future, we will conduct controlled experiments and interviews with social media users to understand how they perceive counter-rumor information and how it changes their behavior and attitudes. 

\section{Conclusion}
In this paper, we integrate quantitative and visual analyses to examine online discussions and debunking activities regarding health rumors on Weibo during the first four months of the COVID-19 crisis. Our study found that two types of health rumors (\ie dread and wish rumors) differed significantly in content and dissemination. We also distinguished between the different behaviors and impacts of four kinds of social media users, from ordinary individuals to government agencies, in combating health rumors. 
We demonstrated the effectiveness of debunking.
Our findings contribute to a better understanding of how online health rumors spread and how they are influenced by debunking efforts in public health crises.
%

%

\section{Acknowledgments}

We are grateful to the anonymous reviewers for their insightful suggestions. We thank Qingbo big data for providing the dataset for this project. We also thank Meng Xia for her great support.
This work is partially supported by the HKUST-SJTU Joint Research Collaboration Fund under Grant No. SJTU20EG02.


\bibliography{aaai22}
\end{document}